\documentclass{article}
\usepackage{graphicx}

\usepackage{color}

\ifx\pdfoutput\@undefined\usepackage[usenames,dvips]{xcolor}
\else\usepackage[usenames,dvipsnames]{xcolor}
\IfFileExists{pdfcolmk.sty}{\usepackage{pdfcolmk}}{}
\fi
\usepackage[plainpages=false,pdfpagelabels,pagebackref=false,naturalnames=true,hyperindex=true,pdftitle={Levels of Abstraction and the Philosophical Legacy of Computability and Information Theory},pdfauthor={Hector Zenil}]{hyperref}
\hypersetup{colorlinks=true,urlcolor=Cerulean,linkcolor=BrickRed,citecolor=RoyalBlue,a4paper,
 pdfpagemode=None,
 pdfstartview=FitH}
\usepackage[all]{hypcap}

\begin{document}

                    
\title{Levels of Abstraction and the Apparent Contradictory Philosophical Legacy of Turing and Shannon}
\author{Hector Zenil\\
\mbox{}\\
Institute d'Histoire et de Philosophie des Sciences\\(Paris 1 Panth\'eon-Sorbonne, CNRS, ENS Ulm), Paris, France; and
\\Unit of Computational Medicine, Karolinska Institute\\ Centre for Molecular Medicine, Stockholm, Sweden.\\ hector.zenil@algorithmicnaturelab.org}

\date{}
\maketitle

\begin{abstract}
In a recent article, Luciano Floridi explains his view of Turing's legacy in connection to the philosophy of information. I will very briefly survey one of Turing's other contributions to the philosophy of information and computation, including similarities to Shannon's own methodological approach to information through communication, showing how crucial they are and have been as methodological strategies to understanding key aspects of these concepts. While Floridi's concept of Levels of Abstraction is related to the novel methodology of Turing's imitation game for tackling the question of machine intelligence, Turing's other main contribution to the philosophy of information runs contrary to it. Indeed, the seminal concept of computation universality strongly suggests the deletion of fundamental differences among seemingly different levels of description. How might we reconcile these apparently contradictory contributions? I will argue that Turing's contribution should prompt us to plot some directions for a philosophy of information and computation, one that closely parallels the most important developments in computer science, one that understands the profound implications of the works of Turing, Shannon and others.
\end{abstract}

Floridi's recent article~\cite{floridi} seems to leave little doubt that Turing's most important contribution to the philosophy of information was the imitation game that he put forward~\cite{turingmind} as a strategy for inquiring into and evaluating the intelligence capabilities of computing machines (today we call it the Turing test):

\begin{quotation}
When one looks at Turing's philosophical legacy, there seems to be two risks. One is to reduce it to his famous test (Turing 1950). This has the advantage of being clear cut. Anybody can recognize the contribution in question and place it within the relevant debate on the philosophy of artificial intelligence. The other risk is to dilute it down into an all-embracing narrative, making Turing's ideas the seeds of anything we do and know today.
\end{quotation}
Reads the Introduction~\cite{floridi}.

One main contribution of Turing's imitation game is methodological in nature, constituting a powerful epistemological approach to under-defined concepts. As Floridi asserts, Turing finds it more appropriate to ask a specific question at the right level of description that can be quantified rather than discussed ad infinitum\footnote{Here it is also worth clarifying the point that his prediction as to when machines would pass the test was rectified by Turing himself in a 1951 BBC radio interview (broadcast a year later in 1952) with H. A. Newman, Sir Geoffrey Jefferson, and R. B. Braithwaite: ``Can Automatic Calculating Machines Be Said to Think?"~\cite{turingradio}, when he offered a prediction of  ``at least 100 years'' (at a 70\% percent chance). So it shouldn't be claimed that he was wrong, as is claimed in \cite{floridi} and many other sources.}.

\section{Meaning and levels of abstraction in Godel's relativization}

As it is well known, at an international mathematics conference in 1928, David Hilbert and Wilhelm Ackermann suggested the possibility that a mechanical process could be devised that was capable of proving all mathematical assertions, this notion referred to as the \emph{Entscheidungsproblem}, or `the decision problem', made not difficult to imagine that arithmetic could be amenable to a sort of mechanisation. The origin of the Entscheidungsproblem dates back to Gottfried Leibniz, who having succeeded (circa 1672) in building a machine, based on the ideas of Blaise Pascal, that was capable of performing arithmetical operations (the \emph{Staffelwalze} or the Step Reckoner), imagined a machine of the same kind that would be capable of manipulating symbols to determine the truth value of mathematical principles. Leibniz devoted himself to conceiving a formal universal language, which he designated `characteristica universalis', a language which would encompass, among other things, binary language and the definition of binary arithmetic.

In 1931, Kurt G\"{o}del arrived at the conclusion that Hilbert's intention (also referred to as `Hilbert's programme') of proving all theorems by mechanizing mathematics was not possible under certain reasonable assumptions. G\"{o}del advanced a formula that codified an arithmetical truth in arithmetical terms and that could not be proved without arriving at a contradiction. Even worse, it implied that there was no set of axioms that contained arithmetic free of true formulae that could not be proved.

Theorems in a mathematical theory are formal semantic objects. They have truth value, conveying information attesting to the truth of the facts encompassed, all the way from axioms, which are facts taken to be true by definition, to the statement of the theorem itself. But Godel did something remarkable and encoded the meaning of theorems in the syntax of the theory itself. He did this by associating symbols with numbers in order to encode meaning in the form of arithmetical propositions. Using a clever construction that led to a contradiction, he proved that some of these constructions are undecidable, that is, they cannot be assigned a meaning within the theory unless a larger more powerful theory is used, which in turn would have new undecidables itself, hence leading to questions of absolute undecidability. 

This fundamental relativisation put an end to the discussion of the feasibility of Hilbert's programme, given that no matter how strong a theory could be, there would always be meaningful statements from outside it that the theory would be unable to encompass. The relationship between truth and the provable was broken.

Just a few years after G\"{o}del, Turing arrived at very similar conclusions by very different means. His means were mechanical, so the theorems and truths from G\"{o}del's work were now nothing but the manipulation of symbols, sequences of tasks as mundane as those which people, then as now, were used to dealing with on an everyday basis, these were computer programs. Turing also showed that no matter how powerful you think a digital computer would be, it would turn out to have serious limitations, notwithstanding its remarkable properties. 

\section{Turing's one machine for everything}

Alan Turing tackled the problem of decision in a different way to G\"{o}del. His approach included the first abstract description of the digital general-purpose computer as we know it today. Turing defined what in his article he termed an `a' computer (for `automatic'), now known as a Turing machine. Turing also showed that certain computer programs can be decided with more powerful computing machines. Unfortunately in the scheme of Floridi's Levels of Abstraction (LoA)~\cite{floridi}, Turing's derivation of a rich hierarchy pertains to incomputable objects. And intermediate degrees of computation are all but natural; examples are non-constructive~\cite{sutner}, hence of little significance to LoA.

As is widely known, a Turing machine is an abstract device which reads or writes symbols on a tape one at a time, and can change its operation according to what it reads, moving forwards or backwards through the tape. The machine stops when it reaches a certain configuration (a combination of what it reads and its internal state). It is said that a Turing machine produces an output if the machine halts, while the locations on the tape the machine has visited represent the output produced.

The most remarkable idea advanced by Turing is his concept of universality, his proof that there is an `a' machine that is able to read other `a' machines and behave as they would for any input. In other words, Turing proved that it was not necessary to build a new machine for each different task; a single machine that could be reprogrammed sufficed for all. Not only does this erase the distinction between programs carried out by different machines (since one machine suffices), but also between programs and data, as one can always codify data as a program to be executed by another Turing machine and vice versa, just as one can always build a universal machine to execute any program.

Turing also proved that there are Turing machines that never halt, and if a Turing machine is to be universal and hence able to simulate any other Turing machine or computer program, it is actually expected that it will never halt for a(n infinite) number of inputs of a certain type (while halting for an infinite number of inputs too). This is something we are faced with in everyday life, for even the simplest and most mundane tasks, approached using devices as simple as Turing machines, already impose limits on our knowledge of these devices and what they are or are not able to compute.

\section{The Shannon legacy}

Once approached the problem of defining an \emph{algorithm} with the concept of Turing computation, a question to be considered concerns the nature of information. Shannon did something similar than Turing for the concept of algorithm, but for the concept of information. Not so long ago the problem of communicating a message was believed to be related only to the type of message and how fast one could send letters through a communication medium. When Morse code was invented, it was clear that the number of symbols was irrelevant, it required only two different symbols to convey any letter and therefore any possible word and any possible message. Shannon separated information from meaning when it came to measuring certain aspects of messages, because meaning seemed to be irrelevant to the question of communication. The same medium could be used for what were thought to be completely different kinds of information, such as images, sounds and text, which today's computers show are not essentially different, being exactly the same at the machine level.

The computer casts the information in a form that we recognize as an image or a sound, a password or an emoticon, but there is no essential difference among these at the level of Shannon information theory. Shannon formally proved~\cite{shannon} that any language, no matter how sophisticated, can be reduced to a 2-symbol system of yes-no answers, and it can be so reduced quite efficiently. This is why we can now store any sort of information in the same device. As Turing showed with respect to computation, information storage too does not require different media; a single medium suffices, indeed any medium (of the same capacity) would suffice albeit noise and other relevant considerations that Shannon himself also studied in incredible formal detail~\cite{shannon}.

That information can be of very different kinds is significant, but what is more remarkable is that all information can be fundamentally treated of the same type, and only the way in which its elements are arranged results in so many disparate meanings, to which Shannon's measures are immune. And if one wished to use Shannon's entropy to distinguish between messages using the same alphabet, one would soon find that it is unsuitable for capturing this level of meaning, as it has been widely recognized starting by Shannon himself. But this is not to say that no formal low-level quantification theory can deal with information and meaning at given levels of abstraction. 

\section{Building on Shannon and Turing}

Algorithmic information theory~\cite{kolmo,chaitin,solomonoff,levin} (AIT), for example, is better at dealing with subtle differences in messages and thereby capturing certain aspects of meaning~\cite{zenilfiloytec}. Its central measure, Kolmogorov complexity ($K$), not only takes into consideration the message itself, but also its recipient and generator. It tells us that a message can be quantified by the length in bits of the shortest computer program that generates it. The computer program reproduces the message and is included with the message itself, so it is in some sense autoexecutable, regardless of the carrier. Barry Cooper points out~\cite{cooper} that

\begin{quotation}
..., if one [limited] oneself to the usual computability models, the notion of randomness of finite strings seems to provide a first step toward a much needed theory of the incomputability of finite objects.
\end{quotation}

The theory of algorithmic information promises to allow some hypothesis testing on the \emph{algorithmicity} of the world~\cite{zenilalgo,zenilfqxi}, and it even introduces the need for an observer~\cite{zenilfiloytec}, given that one cannot calculate $K$ directly but only by indirect methods, making approximations subject to differences in the methods used. But once again, it is not this relativity that makes the theory incredibly powerful, but the objective properties of this quantification of messages and meaning, the fact that the theory provides an invariance theorem asserting that quantifying information content asymptotically converges to the same values regardless of the production method, even though different observers may see different things in the same bit sequence, interpreting it differently.

In~\cite{benthem}, for example, the only reference to AIT as a formal context for the discussion of information content and meaning is a negative one---appearing in van Benthem's contribution (p. 171~\cite{benthem}). It reads:

\begin{quotation}
To me, the idea that one can measure information one-dimensionally in terms of a number of bits, or some other measure, seems patently absurd...
\end{quotation}

I think this position is misguided. When Descartes transformed the notion of space into an infinite set of ordered numbers (coordinates), he did not strip the discussion and study of space of any of its interest. On the contrary, he advanced and expanded the philosophical discussion to encompass concepts such as dimension
and curvature---which would not have been possible without the Cartesian intervention. Perhaps this answers
the question that van Benthem poses immediately after the above remark (p. 171~\cite{benthem}):

\begin{quotation}
But in reality, this quantitative approach is spectacularly more successful, often much more so than anything produced in my world of logic and semantics. Why?
\end{quotation}

Accepting a formal framework such as algorithmic complexity for information content does not mean that the philosophical discussion of information will be reduced to a discussion of the numbers involved---as it did not in the case of the philosophy of geometry or space-time after Descartes. Thanks to Descartes, however, Euclidian geometry eventually exhausted itself, and much of the philosophical discussion was considered complete and settled. But we still pursue the philosophy of Euclidian geometry because we now have a modern perspective that keeps giving us new material with which to approach what was done, how and why, from an hermeneutical perspective. And we have extended the reach of the philosophy of geometry to the philosophy of modern physics. In the future the same will happen for information if we embrace the most recent development in theoretical computer science, with the help of theories such as algorithmic information theory.

\section{Back to Levels of Abstraction}

Some things are more remarkable not because they are different but because they are the same, even if they can be studied in different ways and at different scales and levels of abstraction. Levels of abstraction are necessary for practical reasons. For example, we are used to seeing things at the level at which our physics and biology predispose us to see them; we are finite beings that can store information in certain limited---though extraordinary---ways. One cannot expect to reconstruct information from the bottom up with limited storage capacity and limited understanding. We will never be able to read machine code and see that it is obviously the source code of a sophisticated word processor, even when the machine code is only a plain translation of the computer code in which the software was originally programmed. It is Turing and Shannon who taught us that software in machine code is the same thing as the same software we interact with on our screens.

It is not that it cannot be provided a machine with the step-by-step proof of a mathematical theorem. It's that no one is able to follow such a detailed description before becoming completely lost. In order for information to be useful, it needs to be packaged for human understanding at the right level, that is. Indeed that's why mathematicians have been so good at creating a language for themselves.

Indeed, access and the study of different levels of abstraction are key to understanding our world. Concerned by the return to asking basic questions of the kind considered by Alan Turing within the framework of computability theory, Barry Cooper argues that uncomputability arises at certain levels of causal explanation, at the point of interaction of local and global phenomena~\cite{cooper}, while at another level a phenomenon may be computable~\cite{nature}:

\begin{quotation}
Even in non-linear systems, such high order behaviour [emergent phenomena] is causal --- one phenomenon triggers another. Levels of explanation, from the quantum to the macroscopic, can be applied. But modelling the evolution of the higher-order effects is difficult in anything other than a broad-brush way. Such problems infiltrate all our models of the natural world.
\end{quotation}

Unlike Cooper, I do not think this is an impediment by principle but a practical limitation. But everything else in Cooper's reasoning applies.

It then turns out that Turing's main contribution to computation complements the LoA approach at the end but for different and less fundamental reasons. If in dealing with emergent phenomena, a common task is to identify useful descriptions and to extract enough computational content to enable predictions to be made, then it is clear that one cannot look at natural phenomena at some arbitrary level; one will be able to compute very little if one is trying to extract a biological discovery from a quantum effect. At some level of abstraction, where epistemological limits are of less fundamental nature, the need of LoAs is a pragmatic necessity. Turing's contribution is twofold, on the one hand the novel strategy epitomized by the Turing test suggesting different levels of description and, on the other hand, the seminal concept of Turing universality collapsing levels in a fundamental way. They will appear contradictory if the beauty of their elegant complementarity, fundamental and pragmatical sides, is overlooked.

\end{document}